\documentclass[journal]{IEEEtran}

\DeclareUnicodeCharacter{03A4}{T}

\ifCLASSINFOpdf
\else
\fi
\usepackage{amsfonts}\usepackage{algorithmic}
\usepackage[ruled,vlined]{algorithm2e}

\usepackage{graphicx}
\usepackage{caption}
\usepackage[parfill]{parskip}
\usepackage{amsmath}
\usepackage{subcaption} 
\usepackage{placeins} 

\usepackage{booktabs} 
\usepackage{siunitx} 
\usepackage{makecell} 

\usepackage{booktabs}
\usepackage{adjustbox}
\usepackage{multirow}
\usepackage[caption=false,font=normalsize,labelfont=sf,textfont=sf]{subfig}
\usepackage{stfloats}
\usepackage{url}
\usepackage{verbatim}
\usepackage{cite}

\usepackage{color,soul}

\usepackage{subcaption}

\begin{document}
%
\title{RF-Fencing: A Novel RIS-Based Service for Proactive Covert Communications}

\author{\IEEEauthorblockN{
Alexandros I. Papadopoulos,
Dimitrios Tyrovolas~\IEEEmembership{Member,~IEEE},
Alexandros Pitilakis~\IEEEmembership{Senior Member,~IEEE},
Panagiotis D. Diamantoulakis~\IEEEmembership{Senior Member,~IEEE}, 
Antonios Lalas, 
Konstantinos Votis, \\
Nikolaos V. Kantartzis,~\IEEEmembership{Senior Member,~IEEE},
Sotiris Ioannidis and Christos Liaskos 
}                                      

\thanks{A. Papadopoulos is with the Computer Science Engineering Department, University of Ioannina, Ioannina, Greece, and with the Information Technologies Institute, Centre for Research and Technology Hellas (CERTH), Greece (alexpap@iti.gr, a.papadopoulos@uoi.gr)}
\thanks{D. Tyrovolas, A. Pitilakis, P. D. Diamantoulakis, and N. V. Kantartzis are with the Aristotle University of Thessaloniki, 54124 Thessaloniki, Greece (tyrovolas@auth.gr, alexpiti@auth.gr, padiaman@auth.gr, kant@auth.gr)}
\thanks{A. Lalas, and K. Votis are with the Information Technologies Institute, Centre for Research and Technology Hellas (CERTH), Greece (lalas@iti.gr, kvotis@iti.gr)}
\thanks{S. Ioannidis is with the Foundation for Research and Technology Hellas (FORTH), Greece (sotiris@ece.tuc.gr)}
\thanks{C. K. Liaskos is with the Computer Science Engineering Department, University of Ioannina, Ioannina, and Foundation for Research and Technology Hellas (FORTH), Greece (cliaskos@ics.forth.gr).}

\thanks{This work has received funding from the SNS-JU under the European Union’s Horizon Europe research and innovation programme, in the frame of the NATWORK project (No 101139285) and CYBERSECDOME (No 101120779). The methodology was developed within CYBERSECDOME while NATWORK contributed to the experiments, key results, and conclusions.}
}

\maketitle

\begin{abstract}
Programmable wireless environments (PWEs), empowered by reconfigurable intelligent surfaces (RISes), have emerged as a transformative paradigm for next-generation networks, enabling deterministic control over electromagnetic (EM) propagation to enhance both performance and security. In this work, we introduce RF-Fencing, a novel RIS-enabled PWE service that enforces spatially selective control over wireless transmissions, simultaneously suppressing unwanted signal exposure while sustaining robust connectivity for legitimate users. To realize this vision, we develop SHIELD, a lightweight and scalable algorithm that orchestrates multiple RIS units by multiplexing precompiled codebook entries with real-time, low-complexity optimization. Through extensive evaluations across diverse frequencies, RIS configurations, and deployment scenarios, SHIELD demonstrates both far-field directional control and near-field quiet-zone creation, thereby enhancing network security. Our findings reveal that SHIELD effectively balances proactive covert communication with service delivery by dynamically managing multiple signal suppression and delivery areas, while enabling the realization of EM quiet zones with minimal impact on surrounding regions, ultimately establishing RF-Fencing as a practical RIS-based foundation for privacy-preserving and adaptive wireless environments in future 6G networks.

 \end{abstract}


%
\IEEEpeerreviewmaketitle

\section{Introduction}

The emergence of 6G networks is driving the evolution of wireless systems into multifunctional platforms that support extremely high data rates, integrated sensing and communications, advanced location-based applications~\cite{6GKaragiannidis}, and autonomous mobility~\cite{segata2024cooperis}. However, satisfying these diverse services not only demands ultra reliability but also calls for stringent privacy guarantees to protect sensitive information and maintain system integrity. In this context, covert communications have emerged as a critical paradigm, as they focus on concealing the existence of a transmission, thereby preventing adversaries from even detecting that communication is taking place~\cite{Covert6G}. This approach goes beyond traditional security measures that safeguard information content by ensuring that the act of communication itself remains hidden. Consequently, improving covertness and privacy is now a core 6G goal, enabling next-generation networks to deliver high performance alongside secure and increasingly proactive services against potential threats.

Meeting these diverse requirements further demands network architectures that can be reconfigured with exceptional precision and agility, ensuring high quality of service (QoS) while maintaining strong privacy guarantees. In response to this need, the concept of programmable wireless environments (PWEs) has emerged, where the wireless propagation space is coated with reconfigurable components that actively control electromagnetic (EM) behavior, transforming a traditionally stochastic phenomenon like wireless propagation into a nearly-deterministic one~\cite{liaskos2018new}. Central to PWEs are reconfigurable intelligent surfaces (RISes), which stand out due to their fine-grained control over EM waves and planar design that enables easy deployment across existing infrastructure~\cite{LiaskosProceedings2022}. Specifically, RIS is constructed from periodic sub-wavelength metamaterial elements that are interconnected with active components such as PIN diodes~\cite{10501453,pitilakis2022multifunctional}. Through the optimal configuration of its elements, RIS can dynamically modulate incident EM waves in real time, enabling macroscopic functionalities such as beam steering for targeted signal delivery or absorption for interference suppression and leakage minimization~\cite{open_platform, zeris,pitilakis2023mobility}. As a result, RISes are poised to play a pivotal role in 6G communications by fulfilling strict performance requirements and enabling innovative services that natively support privacy and proactive control of wireless exposure.

Drawing from the vision of PWEs, RIS-enabled covert communications now serve as a compelling enabler for undetectable transmissions in 6G networks~\cite{Covert6G,PLSRIS2024,zerissecurity2025}. Specifically, in contrast to conventional methods that rely on external noise injection or channel uncertainty, RIS can deterministically shape the communication environment, providing fine-grained spatial control of signal propagation~\cite{LiaskosSecurity2019,Naeem2023}. This inherent programmability not only enables reactive mitigation of detected threats, but also opens the door to proactive covert communication, allowing networks to predefine suppression areas and steer signals away from uncertain regions before any adversarial activity arises \cite{shield_pimrc}. As a result, this level of programmability defines a new design space in which covert transmission is not merely layered on top of the physical infrastructure, but is an intrinsic characteristic of the environment itself. Realizing this vision, however, demands the coordinated orchestration of multiple RIS functionalities, where beam steering and spatial suppression are jointly configured to dynamically generate quiet zones and conceal signal paths~\cite{LiaskosSecurity2019,Naeem2023}. To this end, the systematic exploration of RIS-enabled proactive covert communication strategies remains an open and promising research direction, with the potential to fundamentally redefine how privacy and covertness are achieved in wireless networks.

\subsection{State of The Art}

Motivated by the PWE paradigm and the need for privacy-aware networking, recent research has explored the systematic use of RIS for covert communication \cite{Naeem2023, Niyato2020}. Initially, \cite{Niyato2020} demonstrated that RIS configurations can create quiet zones by absorbing or redirecting signals, achieving covert communication with minimal power overhead compared to conventional noise-based techniques. Subsequent works refined this concept; \cite{Basar2021} proposed an alternating optimization strategy jointly tuning RIS configurations and beamforming to maximize secrecy rate under covert constraints, while \cite{Chen2022MIMOCovert} extended the idea to MIMO scenarios using null-space projections for reduced leakage. Nonetheless, these contributions primarily target secrecy performance, neglecting RIS’s potential to manipulate spatial propagation characteristics and effectively suppress adversarial exposure, hence not fully exploiting the capabilities of PWEs.

To further exploit the spatial degrees of freedom offered by RIS, the authors in \cite{Pejoski2022} proposed a full-duplex covert communication scheme where the RIS supports a legitimate downlink transmission to a public user while simultaneously concealing an uplink transmission from a covert user. By jointly optimizing the covert user's power and rate along with the RIS configuration, the approach in \cite{Pejoski2022} revealed the potential of leveraging multi-user interference as a mechanism to preserve covertness without relying on artificial jamming. Similarly, the authors in \cite{Lv2022} adapted this concept to RIS-aided NOMA scenarios, embedding covert transmissions within superimposed downlink signals by exploiting the public user's signal and RIS-induced phase uncertainty. These efforts demonstrated how interference in multi-user contexts can act as a natural masking layer for covert communication. However, neither \cite{Pejoski2022} nor \cite{Lv2022} align with the architectural vision of PWEs, where multiple RIS can be dynamically coordinated to jointly structure the wireless environment \cite{LiaskosProceedings2022,LiaskosSecurity2019}. As a result, these observations underline the need for covert communication strategies that are not only robust, but also capable of leveraging distributed RIS to achieve programmable and spatially-aware covertness.

In response to these limitations, more recent studies have explored expanding RIS-assisted covert communication either by modifying the architecture of individual RIS units or by deploying multiple RIS across the network. Toward the first direction, STAR-RIS has emerged as a promising solution that enables simultaneous reflection and transmission, thereby improving coverage and enhancing propagation control. In more detail, the authors in \cite{Xiao2023STARRIS,Li2024STARRIS} explored the use of STAR-RIS in both multi-antenna and NOMA-based covert scenarios, demonstrating that joint optimization of STAR-RIS configuration and transmission strategies can enhance covert performance by shaping the spatial signal footprint and embedding hidden transmissions within legitimate traffic. In a similar direction, the authors in \cite{Zhu2024ActiveRIS,Kuai2024ActiveNOMA} investigated active RIS architectures with embedded amplification, showing that covert link quality can be improved under stringent detection constraints. However, these approaches do not justify the shift away from nearly-passive designs, as they lack comparison with scenarios where multiple RIS operate cooperatively. In contrast, the authors in \cite{Gao2024MultiRIS} proposed a multi-RIS covert framework based on multi-agent reinforcement learning, showing that coordinated control of distributed RIS can enhance spatial suppression and covert throughput while maintaining the low-power operation expected in PWEs. Nevertheless, the proposed solution remains limited to a single covert link and does not explore the broader potential of multi-RIS coordination to simultaneously support legitimate communication and enforce directional signal suppression as a native feature of the environment. As a result, the architectural potential of multi-RIS systems to realize scalable and programmable privacy-preserving communication remains largely underexplored.

\subsection{Motivation \& Contribution}
Despite the promising progress reported in the literature, most existing approaches have predominantly focused on enhancing covertness, without fully exploiting the software-defined control over EM propagation that RIS can offer. Specifically, many methods rely on heavily iterative optimization frameworks or machine learning-based schemes~\cite{Basar2021, Chen2022MIMOCovert, Pejoski2022, Lv2022}, which pose significant barriers to real-time implementation due to their computational complexity—a challenge common across network security actions~\cite{8272032}. Furthermore, their evaluation remains primarily numerical, lacking integration with physics-based tools that would enhance both accuracy and practical feasibility. Recently, the concept of RIS codebooks has emerged as a compelling solution \cite{nerini2023discrete}, enabling the precompilation of optimized configurations that can be rapidly retrieved during operation~\cite{papadopoulos2024physics}. Building on this, the multiplexing of codebook entries has been proposed to dynamically combine simple functionalities at runtime, allowing the synthesis of complex responses with minimal computational overhead~\cite{Papadopoulos6GNET, segata2024cooperis}. As a result, these developments can be extended to support spatially selective functionalities for private communications, enabling the network to proactively suppress signal exposure in regions of elevated risk where adversaries may be present, even without precise localization, while maintaining scalability and low complexity for the RIS operation~\cite{segata2024cooperis}. To the best of the authors' knowledge, however, no prior work has explored a spatially-aware, codebook-driven framework for covert communication through coordinated multi-RIS control, while simultaneously preserving the QoS for legitimate users.

In this paper, we introduce \textit{RF-Fencing}, a codebook-based service that enables the coordinated control of multiple RIS to realize spatially-aware covert communication through the use of precompiled RIS configurations and real-time multiplexing. Specifically, our contributions are as follows:
\begin{itemize}
    \item We establish the concept of RF-Fencing as a scalable and practical approach for enforcing spatially selective and proactively defined signal suppression within PWEs, leveraging codebook-driven RIS control to create EM quiet zones in regions of elevated risk, while preserving service continuity for legitimate users.
    \item To realize RF-Fencing, \textit{SHIELD} is developed as a dedicated algorithm for multi-RIS coordination, which partitions the network domain into proactively selected hostile signal suppression areas (HSSAs) and friendly signal delivery areas (FSDAs), while also creating fully confined quiet zones dynamically synthesizing the desired EM response by multiplexing precompiled RIS configurations with lightweight optimization.
    \item The effectiveness of SHIELD is validated through comprehensive EM analysis across multiple frequency bands, RIS architectures, and spatial configurations, demonstrating its ability to simultaneously enforce far-field HSSA/FSDA control for directional signal management and near-field quiet zones for volumetric suppression, achieving significant attenuation in targeted regions.
\end{itemize}
\subsection{Structure}
The remainder of the paper is organized as follows. In Section \ref{sec:RIS_modeling}, we present the RF-fencing service and describe how the RIS is modeled via EM-wave analysis in both far and near field case. In Section~\ref{sec:algorithm}, we present the workflow of the proposed SHIELD algorithm. Section~\ref{sec:evaluation} offers the evaluation of the algorithm for different frequency bands, number of HSSAs and FSDAs and quiet zone creation within the network. Finally, Section~\ref{sec:conclusion} concludes the paper.

\begin{figure*}[t]
\centering
\includegraphics[width=\linewidth]{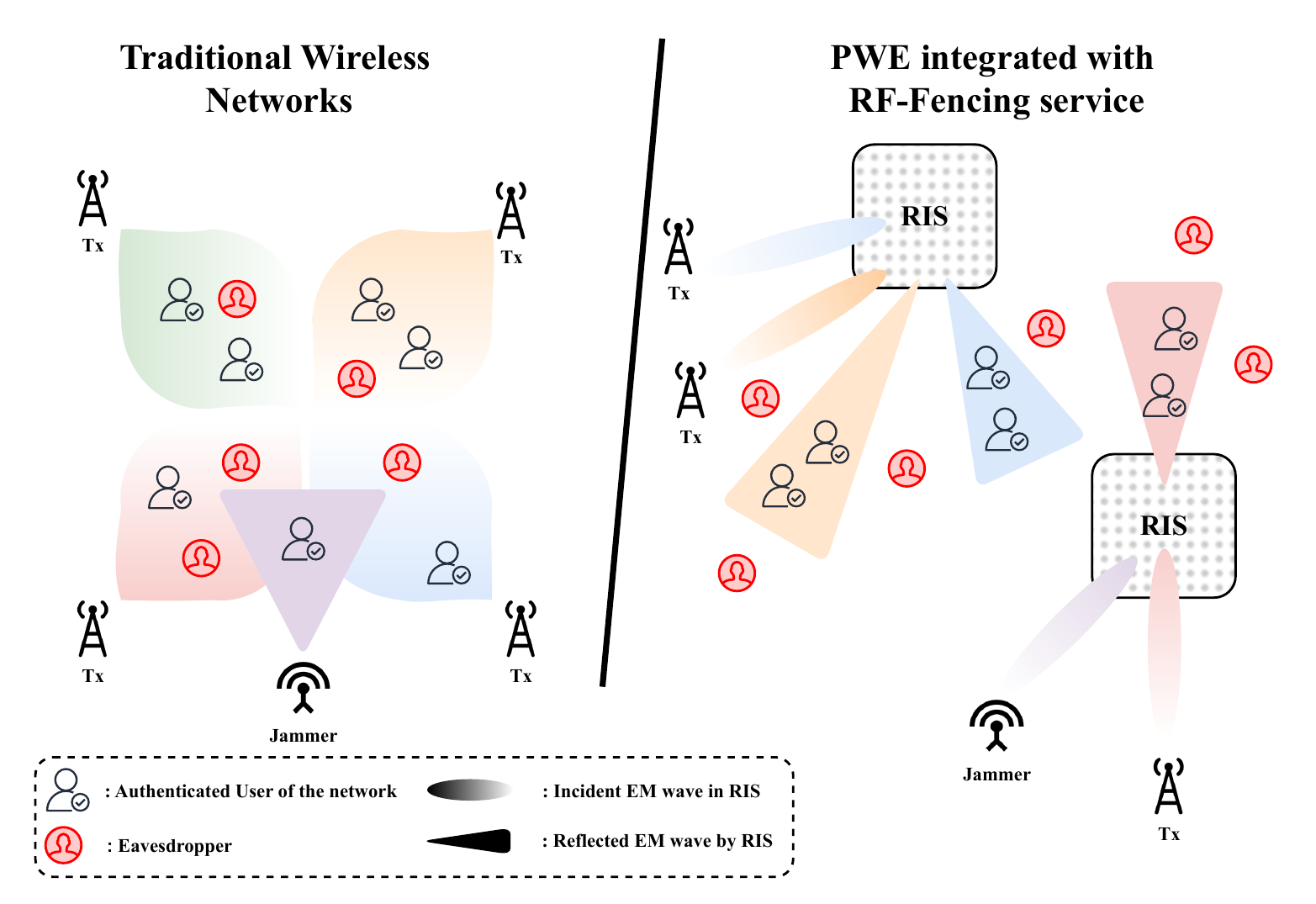}
\caption{Comparison between traditional networks and PWEs in respect of security and covertness }\label{fig:rf_fencing}
\end{figure*}

\section{RF-Fencing \& RIS modeling} \label{sec:RIS_modeling}

As illustrated in Fig. \ref{fig:rf_fencing}, conventional wireless networks allow transmitted signals to propagate freely according to natural propagation phenomena rather than network intent, which leads to unavoidable exposure toward external threats such as jammers and eavesdroppers. To overcome these inherent limitations, there is a growing need for communication environments that can exercise precise control over the propagation of EM waves, shaping signal trajectories according to the network's operational objectives. To this end, PWEs have emerged, integrating multiple RISes that operate as controllable elements within the environment to dynamically manipulate the characteristics of incident signals in real time. Specifically, by coordinating multiple RISes, PWEs can direct wireless signals toward intended receivers while suppressing unwanted transmissions in directions associated with potential adversaries. This capability allows the network to simultaneously enhance communication performance and limit exposure to external threats, thereby laying the foundation for novel PWE-based services that enable covert and private wireless communications by applying spatially selective control over signal propagation.

Building upon this foundation, RF-Fencing addresses the challenge of covert and private wireless communications by enforcing spatially selective signal suppression and delivery through the coordinated control of multiple RISes. Specifically, RF-Fencing leverages the inherent capability of RIS to manage multiple EM functionalities simultaneously, enabling them to receive radiation from various directions and flexibly reroute or cancel it according to network-defined objectives. To efficiently support such diverse functionalities, RF-Fencing adopts the multiplexing of precompiled codebook entries, whereby simpler configurations are dynamically combined to synthesize complex propagation responses~\cite{Papadopoulos6GNET}, effectively eliminating the computational burden associated with iterative optimization. However, as the effectiveness of this approach fundamentally depends on the accuracy of the precompiled configurations, it is essential that they capture the physical behavior of the RIS, particularly for spatially-informed services like covert communications. Consequently, the codebook must be derived from EM-consistent models, since conventional statistical channel abstractions lack the spatial granularity required to ensure reliable covert performance in practical deployments.

To enable the derivation of accurate codebook configurations, physical optics (PO) emerges as a suitable framework for EM-consistent modeling of EM wave propagation, diffraction, and scattering in PWEs, as it offers a tractable yet physics-informed approach to capture the relevant propagation phenomena~\cite{pitilakis2023mobility, Goodman1996}. Unlike statistical methods, PO explicitly models the interactions between EM waves and the surrounding environment, enabling detailed characterization of the suppression and redirection patterns induced by the deployed RIS. Furthermore, PO offers the versatility to accommodate the diversity of propagation scenarios encountered in RF-Fencing, which range from extended network domains to confined physical spaces, by supporting both far-field and near-field modeling perspectives. To this end, in the following subsections, we describe how the PO framework is tailored to accurately model both far-field and near-field propagation conditions, thereby addressing the specific requirements of RF-Fencing across diverse deployment scenarios.


\subsection{Far-Field Modeling}\label{sec:far_field}

In scenarios involving multiple HSSAs and FSDAs distributed across wide areas, the far-field propagation regime becomes applicable, as the RIS-scattered fields are observed at distances significantly larger than the RIS dimensions and the operational wavelength. Under such conditions, the EM behavior of the system is primarily governed by angular distributions, making precise control of propagation directions essential to achieve spatial selectivity. To accurately model these angular characteristics, the Huygens–Fresnel principle (HFP) is adopted, treating the RIS as an ensemble of secondary point sources that re-radiate the incident wavefront toward all directions. In this framework, each RIS is represented as a two-dimensional array composed of $N_{\text{el}} = N_r \times N_c$ reconfigurable elements, also referred to as unit cells. Each unit cell acts as an isolated scatterer, unaffected by its immediate environment, and applies a controllable phase shift to the incident wavefront. Specifically, the controllable phase profile across the RIS is defined as
\begin{equation}
\mathbf{\Phi} = [\Phi_{n}] \in [0, 2\pi]^{N_r \times N_c}, \ n = 1, \dots, N_{el}
\end{equation}
where $\Phi_n$ represents the reflection phase applied by the $n_{th}$ unit. Accordingly, the scattered field produced immediately after interaction with the RIS can be expressed as
\begin{equation}
E_{\text{scat},n} = E_{\text{inc},n} e^{-j \Phi_{n}},
\end{equation}
where $E_{\text{inc},n}$ denotes the incident field at the $n$-th cell, and $\Phi_{n}$ accounts for the imposed phase shift. As described by the HFP, each unit cell acts as a point source emitting a spherical wave, and the superposition of these waves determines the resulting field distribution in any desired direction. This property enables the RIS to dynamically shape its energy distribution over space, allowing the formation of well-defined FSDAs and HSSAs in the far field.

To further detail the system geometry, we consider that the RIS is illuminated by a uniform plane wave of wavelength $\lambda$, arriving from a far-field direction $(\theta_i, \varphi_i)$. Specifically, the RIS elements are arranged on a rectangular grid with uniform spacings $\Delta_x$ and $\Delta_y$ along the horizontal and vertical axes, respectively. Thus, the position of each unit cell is given by
\begin{equation}
x_n = \Delta_x (x - x_c), \quad y_n = \Delta_y (y - n_c),
\end{equation}
where $(x_c, y_c)$ indicates the coordinates of the array center, serving as the reference point for calculating phase differences across the surface. The incident wavefront impinging upon each cell introduces a phase variation that depends on the cell’s position relative to the array center, calculated as
\begin{equation} \label{eq:psi_inc}
\psi_{n}^{\mathrm{inc}} = k \left[ \Delta_x (x - x_c) \cos\varphi_i + \Delta_y (y - y_c) \sin\varphi_i \right] \sin\theta_i,
\end{equation}
where $k = 2\pi / \lambda$ is the wavenumber associated with the operational frequency. By leveraging both the RIS-imposed phase shift and the spatially dependent incident phase, the total contribution from each unit cell is formulated as
\begin{equation}
\Psi_{n} = e ^{j \left( \Phi_{n} + \psi_{n}^{\mathrm{inc}} \right)}.
\end{equation}
Subsequently, the far-field scattered response of the RIS is obtained by summing the contributions of all unit cells, yielding
\begin{equation} \label{eq:e_field}
E(\theta, \varphi) = E_0 \cos^{2\rho}(\theta) \sum_{n=1}^{N_{el}} \Psi_{n},
\end{equation}
where $E_0$ is a normalization constant related to the incident field amplitude (measured in $V/m$), and $\rho$ characterizes the angular selectivity of the scattering pattern. Specifically, $\rho = 0$ corresponds to an isotropic response, while larger values of $\rho$ sharpen the angular focus. To quantify the effectiveness of a given RIS configuration in steering energy toward a specific observation direction $(\theta_d, \varphi_d)$, the field magnitude at the corresponding point of interest (POI) is evaluated as
\begin{equation} \label{eq:E_desired}
E_{\mathrm{POI}} = \left| E(\theta_d, \varphi_d) \right|,
\end{equation}
where $E(\theta_d, \varphi_d)$ represents the scattered field computed using \eqref{eq:e_field} at the desired angles. Finally, it should be noted that under the far-field approximation, these POIs correspond to angular sectors or directional cones, measured in steradians, rather than physical locations within a spatial volume. Consequently, the design of RIS configurations in this regime naturally focuses on controlling the angular radiation characteristics to ensure energy concentration within FSDAs and effective suppression within HSSAs.

\subsection{Near-Field Modeling} \label{sec:near_field}

When considering confined environments, such as the creation of quiet zones within enclosed spaces, the Fresnel–Kirchhoff diffraction (FKD) framework accurately captures intricate wave interactions~\cite{pitilakis2023mobility}. Unlike far-field models that rely on angular propagation, near-field modeling precisely resolves the spatial distribution of EM fields, essential for RF-Fencing deployments targeting localized signal suppression. We evaluate this near-field modeling within a cubic environment of side length $L$. The computational domain employs a dual-grid approach, combining computational efficiency and precise local accuracy. A coarse grid, defined by parameters $N_x$, $N_y$, and $N_z$, uniformly discretizes the entire cubic space $(x,y,z) \in [0,L]^3$. Additionally, a finer grid is constructed specifically within the spherical quiet zone of radius $r_{\text{QZ}}$ centered at $(x_c,y_c,z_c)$, by refining the coarse grid spacing. The refined grid precisely captures EM field variations within the quiet zone while maintaining computational manageability.

The four vertical walls of the space are fully covered by RIS units, each composed of $N_{\text{el}}=N_r \times N_{\text{c}}$ uniformly distributed reflecting elements. These elements are arranged with a margin $m$ from the space boundaries, resulting in a total of $N_{\text{RIS}} = 4 N_{\text{el}}$ scatterers, where the position of the $n$th RIS element is given by $(x_n, y_n, z_n)$. Due to the significant physical dimensions of the RIS surfaces relative to the space size and operating wavelength, the far-field approximation becomes invalid, and a spherical-wave near-field modeling based on FKD is adopted to ensure accurate field estimations within the confined space. Thus, assuming that the source is located inside the room at $(x_s, y_s, z_s)$ and emits an omnidirectional spherical EM wave at frequency $f$, the scattered field within the space is calculated on both grids given by:

\begin{equation}
E_{\text{inc},n} = \frac{E_0}{d_n} e^{j k d_n},
\end{equation}
where $E_0$ represents the amplitude of the source field, and $d_n$ denotes the distance between the source and the $n$th RIS element, which is equal to
\begin{equation}
d_n = \sqrt{(x_n - x_s)^2 + (y_n - y_s)^2 + (z_n - z_s)^2}.
\end{equation}
Each RIS element reflects the incident wave by applying a controllable phase shift $\Phi_n$, while maintaining a unitary reflection magnitude. Consequently, the complex reflection coefficient for the $n$th element is defined as
\begin{equation}
\Gamma_n = e^{j \Phi_n}.
\end{equation}

The contribution of the $n$th RIS element to the scattered field at each observation point is approximated by spherical-wave propagation as
\begin{equation}
E_{\text{scat},n}(x_p, y_p, z_p) = E_{\text{inc},n} \Gamma_n \frac{e^{j k R_n(x_p, y_p, z_p)}}{R_n(x_p, y_p, z_p)},
\label{eq:e_field_nf}
\end{equation}
where $R_n(x_p, y_p, z_p)$ denotes the propagation distance from the $n$th RIS element to the observation point, computed as
\begin{equation}
R_n(x_p, y_p, z_p) = \sqrt{(x_p - x_n)^2 + (y_p - y_n)^2 + (z_p - z_n)^2}.
\end{equation}
The total scattered field $E_{\text{scat}}$ at each grid point is then obtained by summing the contributions from all RIS elements, according to
\begin{equation}
E_{\text{scat}}(x_p, y_p, z_p) = \sum_{n=1}^{N_{\text{RIS}}} E_{\text{scat},n}(x_p, y_p, z_p).
\label{eq:e_scat_room}
\end{equation}
For the purposes of this model, each RIS element is treated as an ideal isotropic point scatterer with individually adjustable reflection phase, while mutual coupling and polarization effects are neglected. Additionally, reflections from uncovered surfaces and secondary reflections among RIS-covered walls are disregarded. The source is assumed to emit a continuous spherical wave at a fixed frequency, maintaining constant amplitude throughout the simulation domain.

Using this modeling framework, we further define a quiet zone in the room as a spatial region where the scattered field is minimized. Specifically, the quiet zone is formally defined as the spherical region satisfying
\begin{equation}
(x_p - x_{\text{c}})^2 + (y_p - y_{\text{c}})^2 + (z_p - z_{\text{c}})^2 \leq r_{\text{QZ}}^2,
\label{eq:qz}
\end{equation}
where $(x_p, y_p, z_p)$ corresponds to the coordinates of each observation point. For the observation points that satisfy  \eqref{eq:qz}, the calculation of  \eqref{eq:e_scat_room} is conducted using the finer and denser computational grid for more accurate results. To quantify the EM field suppression achieved within this region, we evaluate the average magnitude of the scattered field, denoted as $E_{\text{field}}^{\text{avg}}$. This metric directly reflects the residual field energy inside the quiet zone and is computed as
\begin{equation}
E_{\text{field}}^{\text{avg}} = \frac{1}{N_{\text{QZ}}} \sum_{(x_p, y_p, z_p) \in \text{QZ}} |E_{\text{scat}}(x_p, y_p, z_p)|^2,
\end{equation}
where $N_{\text{QZ}}$ denotes the total number of discretized grid points contained within the quiet zone, and $|E_{\text{scat}}(x_p, y_p, z_p)|$ represents the magnitude of the scattered field at each corresponding point. This formulation enables a detailed volumetric analysis of the field distribution within enclosed environments and serves as the foundation for developing RIS control strategies targeting spatially localized suppression objectives, such as EM quiet zones.

\section{SHIELD: A Codebook-Based Algorithm for RF-Fencing}\label{sec:algorithm}

\begin{figure}
\centering
\includegraphics[width=\linewidth]{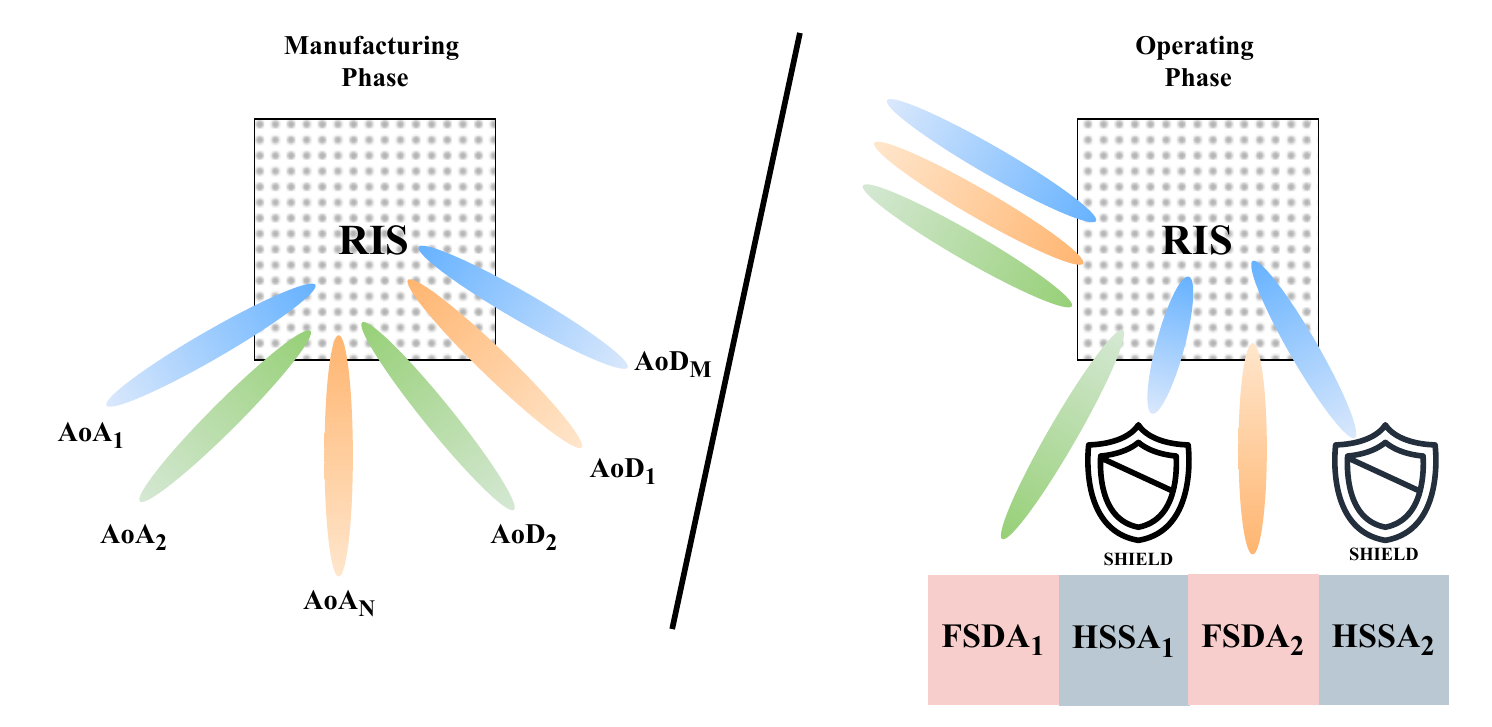}
\caption{Manufacturing (left) and Operating (right) phase using RF-Fencing SHIELD algorithm. }\label{fig:shield_workflow}
\end{figure}

The orchestration of multiple RIS requires a codebook-based control mechanism capable of translating precomputed configurations into real-time operational decisions. In this context, SHIELD is introduced as the first dedicated algorithm designed to enable the practical realization of RF-Fencing within PWEs. Specifically, as illustrated in Fig. \ref{fig:shield_workflow}, SHIELD leverages precompiled codebooks generated during manufacturing, which detail RIS configurations for various predefined beam steering scenarios encompassing different angles of arrival ($\textit{AoA}$) and departure ($\textit{AoD}$) \cite{segata2024cooperis,Papadopoulos6GNET}. Since these configurations are generated during manufacturing, SHIELD simply retrieves and multiplexes the codebook information, then employs a lightweight optimization procedure to generate the RIS configuration matrix $\mathbf{\Phi}$ that suppresses EM radiation in undesired directions while maintaining signal quality within intended coverage zones, thus avoiding computationally intensive iterative optimization and full channel estimation at runtime. Notably, SHIELD can also function as a jamming mitigation mechanism by transferring the area containing legitimate users into HSSAs when the jammer’s $\textit{AoA}$ is known, effectively canceling interference. The overall workflow of SHIELD, encompassing codebook retrieval, field multiplexing, and online phase refinement, is presented in Alg.\ref{alg: SHIELD}.

\begin{algorithm}[t]
\caption{ RF-Fencing Algorithm for Hostile Signal Suppression and Friendly Signal Delivery Areas Creation Within the Network (SHIELD)}
\begin{algorithmic}[1]
\small
\STATE \textbf{Input:} 
\begin{itemize}
    \item Codebook Database with \(E_k\) for supported \textit{AoA}/\textit{AoD}.
    \item Number of $d,u$ for FSDAs and HSSAs.
    \item Tolerance and learning rate $\mu$ and weight parameter $w_{opt}$.
    \item Thresholds \(\tau_{FSDA}\), \(\tau_{HSSA}\).
    \item Compromise factor $\eta$.
\end{itemize}
\STATE \textbf{Dominant Field Identification:} For each case \(k\), identify dominant field values.
\STATE Form the masks for FSDA and HSSA, \(\mathcal{M}_d\) and \(\mathcal{M}_u\), respectively using Eq. \eqref{eq:dominant2}.
\STATE \textbf{Common Field Computation:} Compute \(\mathbf{E_{\text{common}}}\) as in Eq. \eqref{eq:e_common} and obtain the initial $\mathbf{\Phi}_{int}$ from Eq. \eqref{eq:Phi_common}.
\STATE \textbf{Online Optimization:}
    \begin{itemize}
        \item Set \(\mathbf{\Phi}_{\text{init}} \gets \mathbf{\Phi}_{\text{common}}\).
        \item Define the cost function by evaluating Eqs. \eqref{eq:J_d/u},\eqref{eq:J_total}.
        \item Update the $\mathbf{\Phi}$ using a gradient descent step according to Eq. \eqref{eq:optimizatio}:
        \[
        \mathbf{\Phi}_{\text{opt}} = \mathbf{\Phi}_{\text{init}} - \mu \nabla_{\mathbf{\Phi}}\mathcal{J}(\mathbf{\Phi}).
        \]
    \end{itemize}
\STATE \textbf{Output:} Final common phase profile \(\mathbf{\Phi}_{\text{opt}}\) and performance metrics for FSDAs and HSSAs (Eq. \eqref{eq:performance}).
\end{algorithmic}
\label{alg: SHIELD}
\end{algorithm}

The operation of SHIELD begins with the identification of angular regions that correspond to both service provisioning and signal suppression objectives, leveraging precomputed EM field responses $E_{\text{field}}$ obtained from the RIS codebook. To this end, the algorithm takes as input the desired number of FSDAs and HSSAs, denoted by $d$ and $u$, respectively. For each of them, SHIELD analyzes the associated field distribution $\left| E_{\text{field}} \right|$ across the angular domain in order to detect dominant scattering components. In this process, angular regions where the field magnitude exceeds the predefined thresholds $\tau_{\text{FSDA}}$ for FSDAs or $\tau_{\text{HSSA}}$ for HSSAs are systematically aggregated into direction-specific masks, which can mathematically be expressed as

\begin{equation}
\begin{aligned}
\mathcal{M}_d &= \bigcup_{k=1}^{d}  \left\{\,\theta,\varphi \;:\; |E_k| >
        \tau_{\mathrm{FSDA}}\,
        \max |E_d(\theta,\varphi)| \right\} \\[2pt]
\mathcal{M}_u &= \bigcup_{k=d+1}^{d+u} \left\{\,\theta,\varphi \;:\; |E_k| >
        \tau_{\mathrm{HSSA}}\,
        \max |E_d(\theta,\varphi)| \right\}
\end{aligned}
\label{eq:dominant2}
\end{equation}

where $\mathcal{M}_d$ and $\mathcal{M}_u$ represent the sets of angular positions associated with the FSDA and HSSA regions, respectively, and $|E_k|$ corresponds to the field magnitude retrieved from the $k$-th codebook entry. These masks effectively localize the angular sectors relevant for service provisioning and signal suppression.

Building upon the constructed masks, SHIELD proceeds to synthesize a composite EM response $\mathbf{E_{\text{common}}} $ that reinforces the dominant fields within the FSDA regions while simultaneously suppressing emissions toward the identified HSSAs. To achieve this, SHIELD computes the maximum of the FSDA field components in areas where there is no overlap with the HSSAs, while in overlapping regions, it applies a compromise factor $\eta$ to balance the conflicting objectives of suppression and service preservation. Outside these designated regions, the field components are deliberately nulled to minimize unintended radiation. Thus, the composite field can be expressed as 
\begin{equation}
\mathbf{E_{\text{common}}} = 
\begin{cases} 
\displaystyle \max_{1 \le k \le d} E_k 
  & \text{in } \mathcal{M}_d \setminus \mathcal{M}_u, \\[1ex]
\displaystyle \eta\,\max_{1 \le k \le d} E_k 
  & \text{in } \mathcal{M}_d \cap \mathcal{M}_u, \\[1ex]
 & \text{otherwise},
\end{cases}
\label{eq:e_common}
\end{equation}
where $\eta$ serves to control the trade-off between maintaining service quality and enhancing suppression in overlapping regions. Following this, the initial phase configuration for all RIS elements, denoted by $\mathbf{\Phi}_{\text{common}}$, is extracted from the argument of the composite field, as given by 
\begin{equation}
\mathbf{\Phi}_{\text{common}} = \arg(\mathbf{E_{\text{common}}} ),
\label{eq:Phi_common}
\end{equation}
where the operation $\arg(\cdot)$ returns the phase angle of the complex field values. This step provides an efficient near‐optimal estimation of the $\mathbf{\Phi}_{\text{common}}$, similar to methods used in smart antenna arrays and analog beamforming for both far-field and near‐field cases \cite{balanis2016antenna}. Consequently, a lightweight gradient-based procedure is used in order to fine-tune the initial guess of $\mathbf{\Phi}_{\text{common}}$ and correct minor deviations \cite{liaskos2020internet}. By this approach, SHIELD workflow remains applicable for real-time RIS control offering the minimal computational overhead. Specifically for the optimization stage, SHIELD employs a cost function that quantifies the discrepancy between the synthesized EM field and the desired target amplitudes within the angular masks $\mathcal{M}_d$ and $\mathcal{M}_u$, corresponding to the FSDAs and HSSAs, respectively. For each region type, this mismatch is computed as
\begin{equation} 
\mathcal{J}_{d/u}(\mathbf{\Phi}) = \sum_{(\theta,\varphi) \in \mathcal{M}_{d/u}} \left|E(\mathbf{\Phi}; \theta,\varphi) - E_{\text{POI},d/u}\right|^2,
\label{eq:J_d/u} 
\end{equation}
where $E(\mathbf{\Phi}; \theta,\varphi)$ represents the EM field evaluated under the current RIS configuration $\mathbf{\Phi}$, and $E_{\text{POI},d/u}$ denotes the target field amplitude within the FSDAs and HSSAs, respectively. These region-specific costs are then combined into a unified objective function, formulated as
\begin{equation} 
\mathcal{J}(\mathbf{\Phi}) = \mathcal{J}_d(\mathbf{\Phi}) - w_{\text{opt}} \, \mathcal{J}_u(\mathbf{\Phi}),
\label{eq:J_total} 
\end{equation}
where the weight parameter $w_{\text{opt}}$ regulates the balance between maximizing field strength in the FSDAs and suppressing leakage toward the HSSAs. Accordingly, the overall optimization problem is expressed in the standard form as
\begin{equation}
\mathbf{\Phi}_{\text{opt}} = \arg\max_{\mathbf{\Phi}} \, \mathcal{J}(\mathbf{\Phi}),
\label{eq:optimization_problem}
\end{equation}
which is efficiently solved via a single-step gradient descent update, expressed as
\begin{equation} 
\mathbf{\Phi}_{\text{opt}} = \mathbf{\Phi}_{\text{init}} - \mu \nabla_{\mathbf{\Phi}}\mathcal{J}(\mathbf{\Phi}), 
\label{eq:optimizatio} 
\end{equation}
where $\mu$ represents the learning rate (in $\text{m}^2/\text{V}^2$), and the initialization $\mathbf{\Phi}_{\text{init}}$ is set equal to $\mathbf{\Phi}_{\text{common}}$. Therefore, the final optimized phase profile $\mathbf{\Phi}_{\text{opt}}$ is then applied across the RIS elements to realize the spatial signal shaping required for the RF-Fencing service.

To quantify the performance of SHIELD, we utilize $P_k$ which is defined as the ratio of the electric field magnitudes at the POIs between the final optimized configuration and the initial beam-steering scenario, which is given as 
\begin{equation} 
P_k = 20 \log_{10} \left( \frac{E_k^{\text{POI,final}}}{E_k^{\text{POI,initial}}} \right), 
\label{eq:performance} 
\end{equation} 
where $E_k^{\text{POI,final}}$ and $E_k^{\text{POI,initial}}$ denote the electric field magnitudes at the POI after applying SHIELD and under the initial, optimal beam-steering configuration, respectively. In the case of FSDAs, values of $P_k$ approaching 0 dB indicate that users experience nearly identical performance compared to conventional beam steering, thus preserving communication quality. Conversely, for HSSAs, the objective is to maximize suppression, leading to strongly negative values of $P_k$ that reflect effective mitigation of unintended emissions. In scenarios involving multiple FSDAs or HSSAs, the overall performance is obtained by computing the individual amplitude ratios for each region using Eq. \eqref{eq:performance} and averaging the resulting field values across all corresponding POIs.

Finally, the computational complexity of SHIELD is determined by the RIS size \(N_{\text{el}} = N_r \times N_c\) and the number of iterations \(N_I\) required for convergence. Both dominant region identification and multiplexed field computation scale linearly with the number of RIS elements, i.e., \(\mathcal{O}(N_{\text{el}})\), while the gradient-based refinement step requires full-array updates at each iteration, resulting in an overall complexity of \(\mathcal{O}(N_I \cdot N_{\text{el}})\).

\section{Performance Evaluation} \label{sec:evaluation}

In this section, we evaluate the performance of SHIELD in managing both directional beam control and spatial region suppression, addressing distinct propagation regimes and deployment requirements. First, we evaluate SHIELD’s capability to dynamically slice the network into FSDAs and HSSAs, steering and suppressing EM emissions under varied conditions. Following this, the evaluation extends to near-field environments, where the focus shifts from directional beam management to the creation of spatially confined quiet zones within enclosed spaces, ensuring localized EM silence without degrading performance in the surrounding areas. Across all scenarios, the incident plane wave is consistently defined with a reference magnitude of $E_0 = 1~ \text{V/m}$, serving as a baseline for evaluating the behavior of the scattered field.

\subsection{PWE Slicing into FSDAs and HSSAs}

We first focus on the far-field regime, where the control of RIS-scattered fields is inherently determined by angular distributions, and varying operating frequencies and RIS configurations directly influence both the physical system behavior and the complexity of the optimization process. Specifically, by examining a THz scenario and one mmWave scenario, we demonstrate SHIELD’s ability to dynamically manage multiple FSDAs and HSSAs across diverse deployment contexts and propagation conditions.

\subsubsection{THz Scenario}

The performance of SHIELD is first evaluated within a THz-band communication network, operating at a carrier frequency of 1~THz. The RIS is configured as a $50 \times 50$ element array, with an inter-element spacing of $\lambda_w/5$. To model the EM behavior of the system, the far-field expressions, as described in Section \ref{sec:far_field}, are employed, computing the scattered field $E_{\text{field}}$ over an angular resolution of $1^\circ$ to ensure precise evaluation of spatial field variations. During the manufacturing phase, a total of 500 precomputed, optimal beam-steering configurations, covering combinations of ($\textit{AoA},\textit{AoD}$), are stored within the codebook for use during operation. The online evaluation focuses on the creation of two FSDAs and one HSSA. Moreover, the input parameters for SHIELD include a convergence tolerance of $10^{-3}$, an optimization weight factor of $w_{\text{opt}} = 0.5$, a learning rate of $\mu = 0.02$, and a compromise factor $\eta = 0.75$. Finally, the thresholds for identifying dominant regions are set to $\tau_{\text{FSDA}} = 0.95$ and $\tau_{\text{HSSA}} = 0.96$, respectively.

\begin{figure}
\centering
\includegraphics[width=\linewidth]{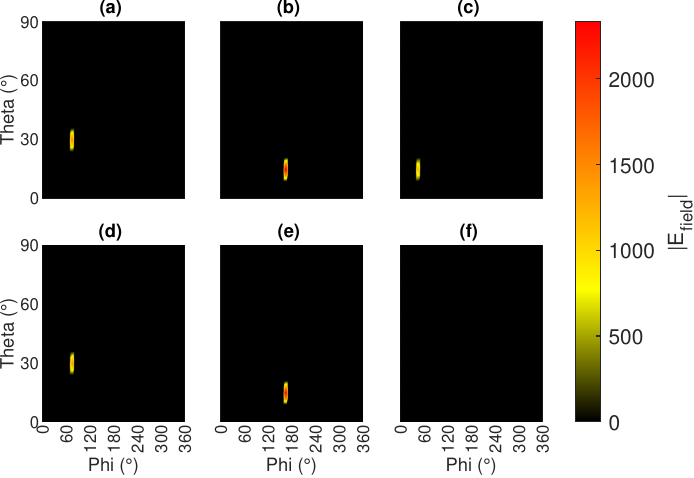}
\caption{$E_\text{field}$ magnitude before (top) and after (bottom) SHIELD usage for FSDAs ($30^\circ$, $75^\circ$), ($15^\circ$, $165^\circ$) and HSSA ($15^\circ$, $45^\circ$).}
\label{fig:example}
\end{figure}

To illustrate SHIELD’s behavior in this scenario, Fig.~\ref{fig:example} presents the $E_{\text{field}}$ distributions before and after optimization. In more detail, the initial beam-steering configurations are shown in the top row, targeting the selected FSDAs at $(30^\circ, 75^\circ)$ and $(15^\circ, 165^\circ)$, alongside the identified HSSA at $(15^\circ, 45^\circ)$. In this initial condition, the $E_{\text{POI}}$ values measured at the two FSDAs are $1530.9\,\text{V/m}$ and $2332.5\,\text{V/m}$, respectively, whereas the HSSA exhibits a field magnitude of $1207.4\,\text{V/m}$. After applying SHIELD, the HSSA field falls to $0.41\,\text{V/m}$—a $-69.25$~dB reduction—while the FSDAs still receive $1447\,\text{V/m}$ and $2269.4\,\text{V/m}$, corresponding to only $-0.49$~dB and $-0.24$~dB losses in respect of their optimal performances. These results collectively illustrate SHIELD's effectiveness in suppressing undesired emissions towards the HSSA while maintaining the signal quality within the intended service regions.

\begin{figure}
\centering
\includegraphics[width=\linewidth]{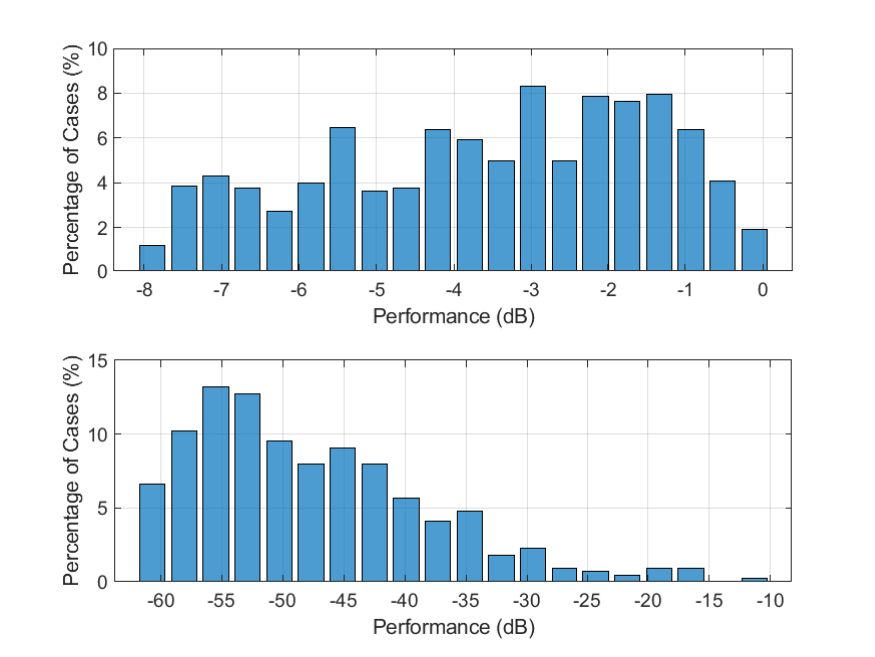}
\caption{Cumulative performance results across multiple cases for FSDAs (up) and HSSAs (down). }
\label{fig:results_percentage_thz}
\end{figure}

Fig.~\ref{fig:results_percentage_thz} provides a comprehensive summary of SHIELD’s performance across 500 evaluated scenarios, highlighting both suppression effectiveness and service continuity. For HSSAs, SHIELD achieves suppression levels exceeding $-50$~dB in approximately $49.1\%$ of cases, between $-50$~dB and $-20$~dB in $49.32\%$, and greater than $-20$~dB in only $1.59\%$. These results confirm SHIELD’s consistent capability to mitigate unwanted emissions across a wide range of spatial configurations. In terms of FSDA, the distribution of performance $E_{\text{POI}}$ illustrates SHIELD’s ability to preserve communication quality for legitimate users despite the stringent suppression requirements. Specifically, about $29.3\%$ of cases experience less than $2$~dB reduction relative to a beam-steering baseline—where the RIS serves each user optimally and independently without simultaneous HSSA suppression. Moreover, $30.7\%$ and $23.9\%$ of cases exhibit moderate reductions in $E_{\text{POI}}$ within the ranges $[-4,-2]$~dB and $[-6,-8]$~dB, respectively, while only $16.1\%$ show more notable degradations up to $-8$~dB.
These results underscore SHIELD’s robustness in balancing precise spatial suppression with reliable service provision, confirming its suitability for enabling covert communications in complex environments with multiple, spatially distinct performance objectives.

\begin{figure}[t]
\centering
\includegraphics[width=\linewidth]{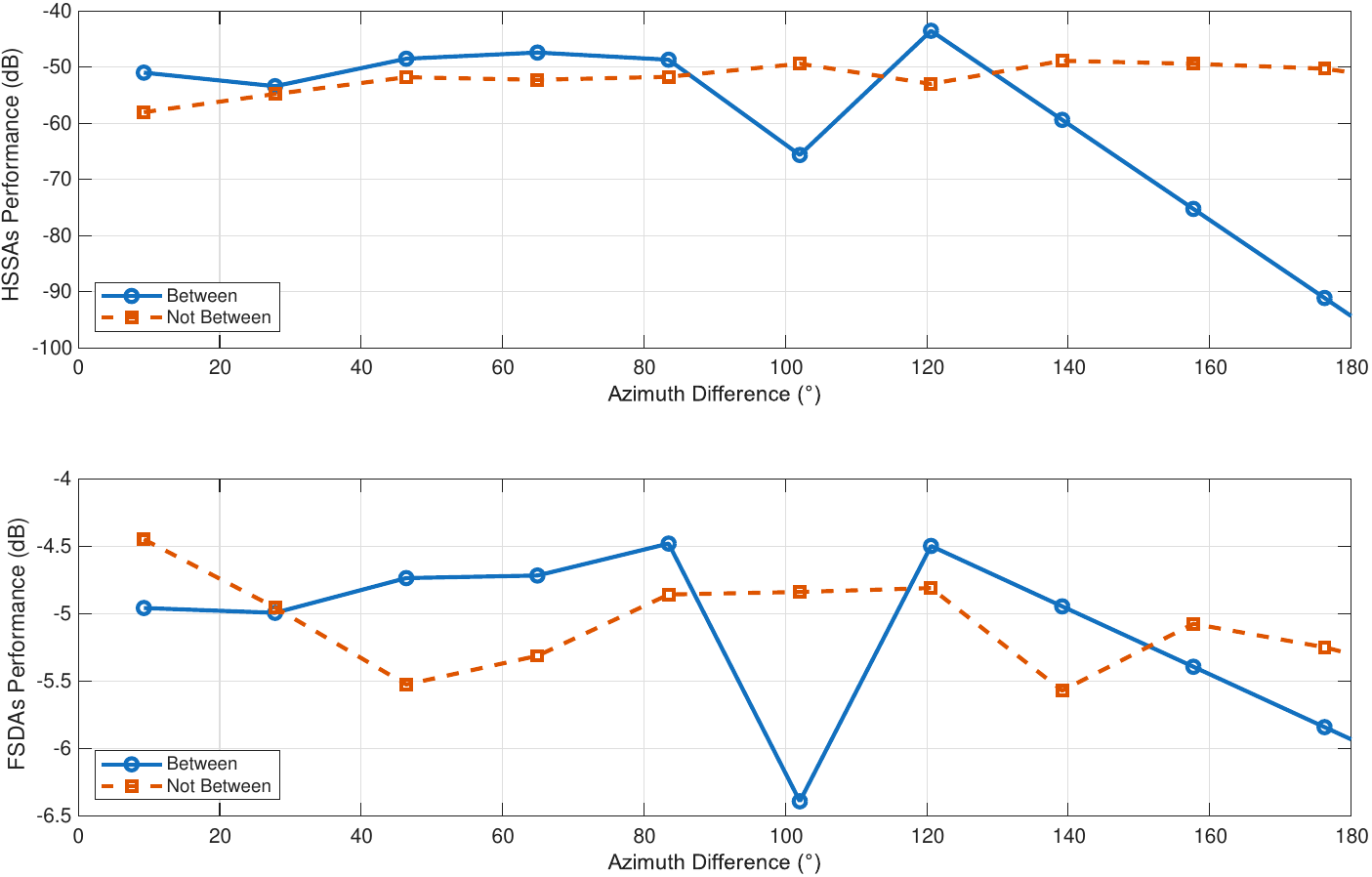}
\caption{Azimuthal difference effect when the single HSSA lies between (blue) or outside (red) the two FSDAs.}
\label{fig:azimuth_thz}
\end{figure}

\begin{figure}[t]
\centering
\includegraphics[width=\linewidth]{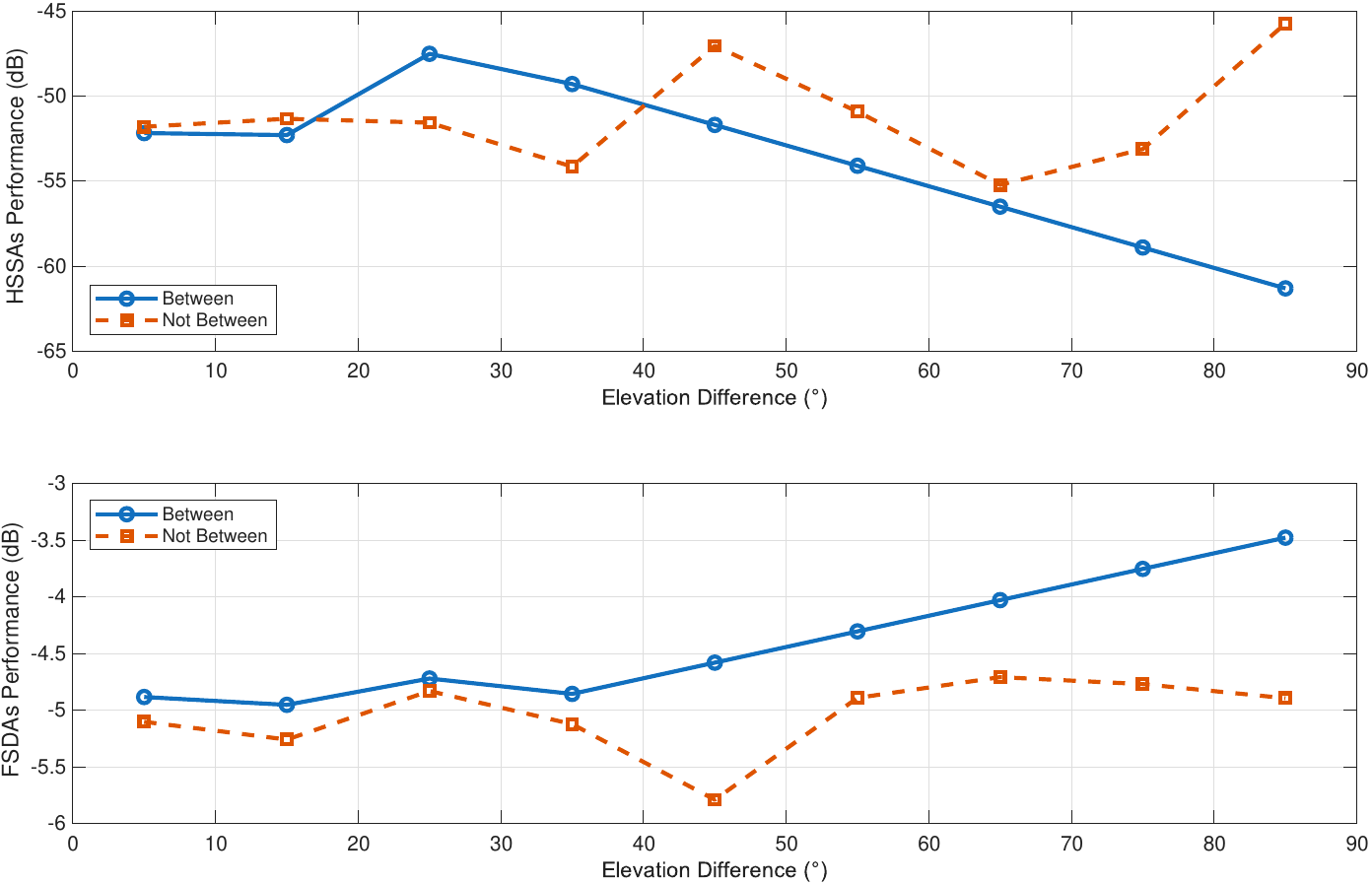}
\caption{Elevation difference impact on SHIELD performance for the same HSSA/FSDA layout categories as in Fig.~\ref{fig:azimuth_thz}.}
\label{fig:elevation_thz}
\end{figure}

Finally, Figs. \ref{fig:azimuth_thz} and \ref{fig:elevation_thz} analyze the influence of angular separation between HSSAs and FSDAs, as well as physical constraints, on SHIELD’s performance. Specifically, in Fig. \ref{fig:azimuth_thz}, performance is plotted against azimuthal separation for two layouts: one with the HSSA outside the angular span of the two FSDAs, and one with the HSSA between them. When the HSSA lies outside the FSDA span, SHIELD consistently achieves suppression of approximately $-50$ to $-60$~dB, with only controllable impact on FSDA performance relative to the unconstrained beam-steering baseline. In the more challenging arrangement, with the HSSA between the FSDAs, SHIELD maintains robust suppression even at narrower angles. When the angular separation becomes clearer, suppression is also achieved up to $-80$~dB. 
Fig.~\ref{fig:elevation_thz} reveals similar trends with elevation angle differences, showing suppression consistently within approximately $-45$ to $-65$~dB for the HSSA, while FSDA performance approaches its optimal value as elevation separation increases. Therefore, Figs. \ref{fig:azimuth_thz} and \ref{fig:elevation_thz} confirm the effectiveness of SHIELD in ensuring substantial suppression and service continuity, validating its applicability in dense multi-RIS deployments.

\subsubsection{mmWave Scenario}

To further evaluate the versatility of SHIELD under distinct propagation conditions, we consider an mmWave communication network operating at a carrier frequency of $300$~GHz. In this setup, the RIS comprises a $75 \times 75$ element array with an inter-element spacing of $\lambda_w/5$. The associated codebook includes $500$ precomputed beam-steering configurations, ensuring adequate coverage across the angular domain despite the broader beam profiles inherent at this frequency compared to the THz scenario. The goal here is the slicing of the network into two HSSAs and one FSDA, trying to estimate the SHIELD performance in a more challenging scenario. The evaluation encompasses $500$ distinct cases. Moreover, SHIELD is applied using for threshold values $\tau_{\text{FSDA}}=0.97$ and $\tau_{\text{HSSA}}=0.99$. The compromise factor $\eta$ is equal to 0.1 while the optimization weight factor is set to $w_{\text{opt}} = 150$. The learning rate $\mu$ is kept equal to 0.02.

\begin{figure}
\centering
\includegraphics[width=\linewidth]{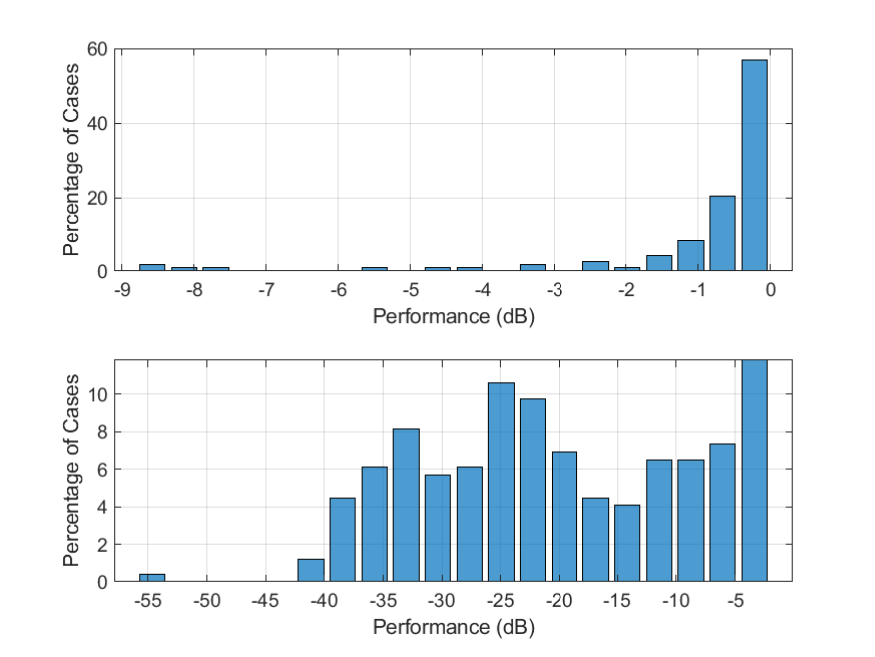}
\caption{Cumulative performance results across multiple cases for FSDAs (up) and HSSAs (down).}
\label{fig:results_percentage_mmwave}
\end{figure}

\begin{figure} [t]
\centering
\includegraphics[width=\linewidth]{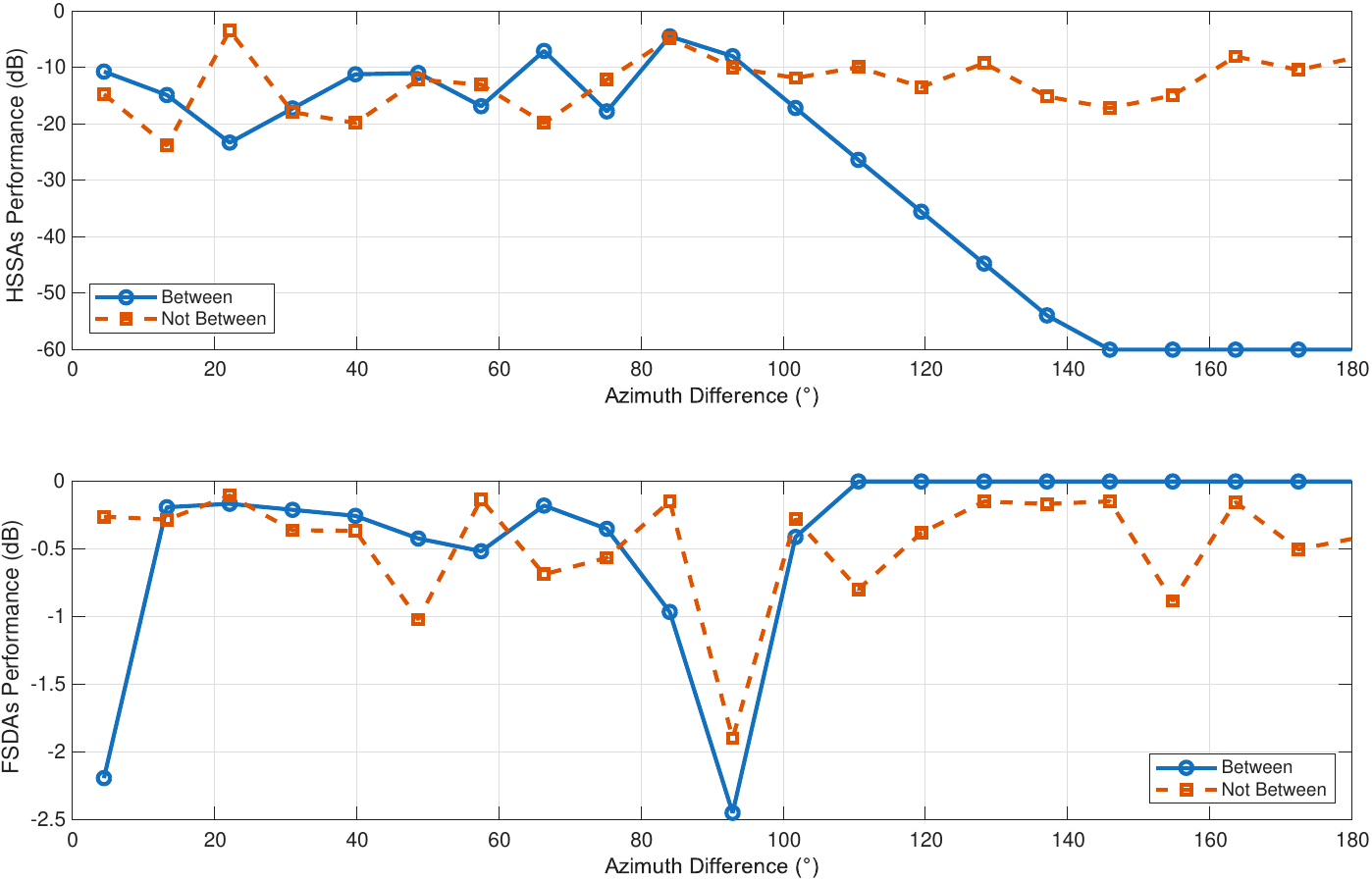}
\caption{Azimuth difference when FSDA is between (blue) and not between (red) the HSSAs.}
\label{fig:azimuth_mmwave}
\end{figure}

Fig.~\ref{fig:results_percentage_mmwave} presents the cumulative performance distribution for all evaluated mmWave cases, highlighting SHIELD’s ability to achieve reliable suppression in undesired directions while maintaining high service quality in intended areas. In the HSSAs, suppression exceeds $-30$~dB in approximately 25\% of cases, with a further 50\% achieving attenuation between $-30$~dB and $-10$~dB, confirming consistent mitigation of unwanted emissions across most scenarios. Among the remaining cases, suppression remains significant, largely within the range of $-10$~dB to $-5$~dB. In terms of FSDA performance, 89.43\% of cases exhibit negligible impact relative to the optimal beam-steering baseline, with 4.8\% and 2.4\% experiencing minor reductions within $[-4,-2]$~dB and $[-6,-4]$~dB, respectively. The final 0.81\% of cases show more substantial reductions beyond $-8$~dB, reflecting the inherent trade-offs involved in managing multiple HSSAs simultaneously. These results demonstrate SHIELD’s capability to balance strong suppression with service continuity, even under the broader beam patterns and complex propagation conditions characteristic of mmWave environments.

Fig.~\ref{fig:azimuth_mmwave} examines the influence of azimuthal separation between HSSAs and FSDAs on the performance of SHIELD in mmWave environments, considering configurations where the FSDA is either located between or outside the angular span of the HSSAs, as before. When the FSDA is placed outside this range, HSSA suppression stays between $-10$~dB to $-20$~dB for most separations, and the FSDA achieves optimal performance or at most a $-2$~dB loss. When the FSDA lies between the HSSAs, HSSA suppression worsens by about  $-1$~dB on average, while the FSDA performance remains near its optimal value. As the azimuthal difference increases, HSSA suppression improves sharply—reaching up to $-60$~dB, while the FSDA’s performance returns to the baseline optimum. Thus, Fig.~\ref{fig:azimuth_mmwave} underscores SHIELD's ability to dynamically adapt to spatial constraints, with performance gains becoming more pronounced as angular diversity increases, highlighting its strategic advantage in managing interference and maintaining selective coverage in complex environments.

\begin{figure*}[t]
  \centering
  \begin{subfigure}[b]{0.33\textwidth}
    \includegraphics[width=\linewidth]{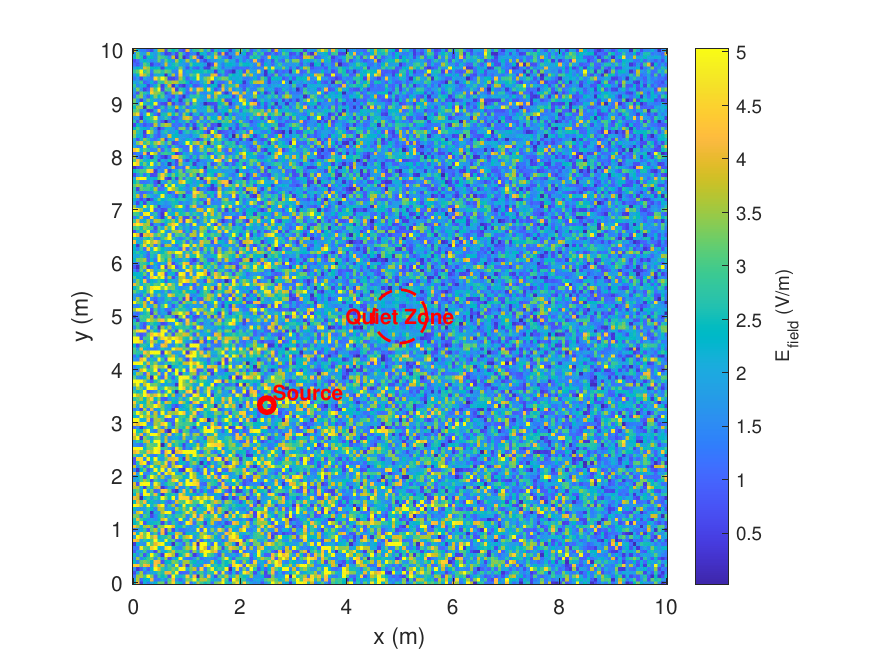}
    \caption{}
    \label{fig:indoor_initial}
  \end{subfigure}\hfill%
  \begin{subfigure}[b]{0.33\textwidth}
    \includegraphics[width=\linewidth]{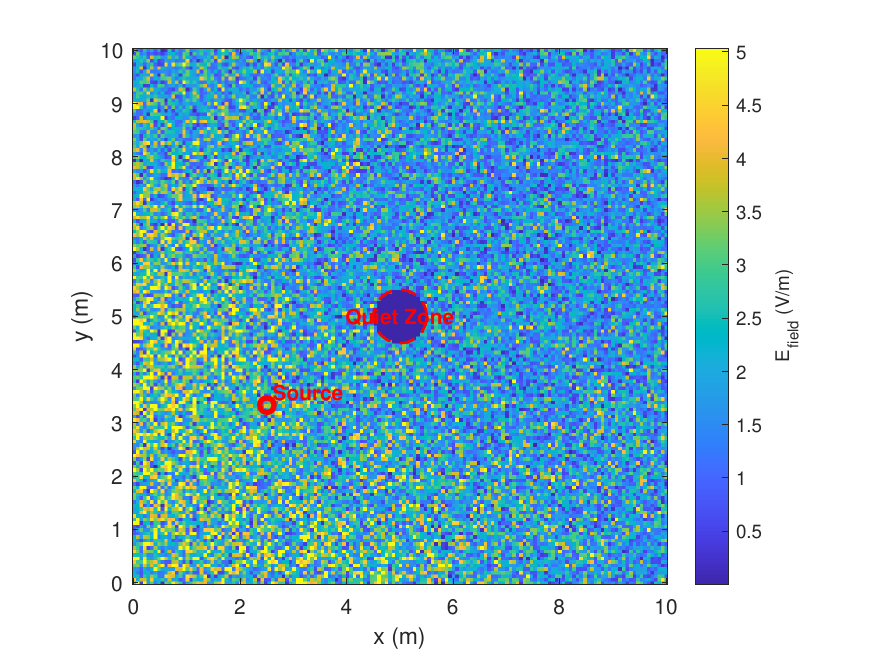}
    \caption{}
    \label{fig:indoor_final}
  \end{subfigure}\hfill%
  \begin{subfigure}[b]{0.33\textwidth}
    \includegraphics[width=\linewidth]{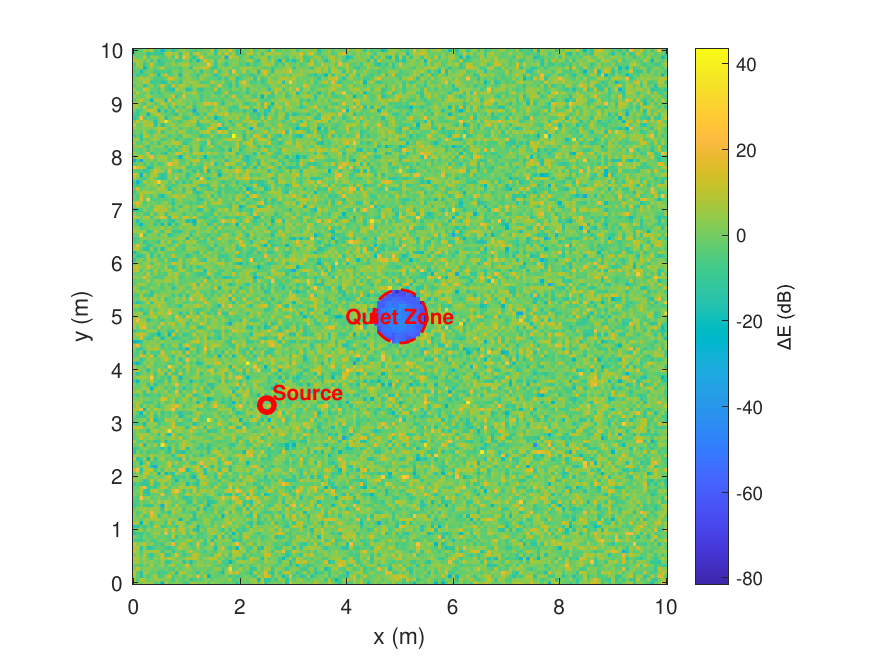}
    \caption{}
    \label{fig:difference_indoor}
  \end{subfigure}
  \caption{Spatial distribution of $E_{\text{field}}$ (in V/m) before (left) and after (center) SHIELD, and the corresponding point-wise deviation (right) in dB.}
  \label{fig:indoor}
\end{figure*}

\subsection{Quiet Zone Creation within PWE}

The next step of our evaluation focuses on the creation of a quiet zone within an enclosed environment, aiming to suppress EM leakage within a specific spatial region while preserving reflections elsewhere. Specifically, in dense indoor settings, the objective is to demonstrate fine-grained EM control for localized suppression, where the confined environment amplifies the impact of multipath reflections and the close proximity of RIS elements intensifies the need for highly precise tuning to effectively isolate the quiet zone. Conversely, in expansive outdoor scenarios, the emphasis moves to validating SHIELD's scalability and its ability to sustain effective suppression across broader areas, where the propagation environment becomes increasingly complex and the trade-off between control precision and coverage must be carefully managed. Thus, by exploring both scenarios, we comprehensively assess SHIELD’s capability for quiet-zone formation across various realistic deployment conditions.

In this direction, SHIELD is applied to jointly optimize four RIS units positioned along the walls of a square area of side length~$L$, covering both enclosed indoor spaces and larger-scale outdoor environments. The objective is to minimize the RIS-scattered electric field $E_{\text{field}}$ within a predefined suppression region of radius~$R_0$ considering it as an HSSA and ensuring proactive covertness against potential wardens in the targeted area. The source transmitter is assumed to be freely positioned at any point within the space, reflecting a realistic deployment scenario and the $E_{\text{field}}$ in each region within the room is computed as detailed in Sec. \ref{sec:near_field}. The process begins by configuring all RIS units as perfect electric conductors, initializing their reflection phases uniformly to~$\Phi_{n} = \pi$. As a first step, SHIELD computes for each RIS the phase shift that minimizes the EM propagation in the quiet zone while keeping all other elements fixed. This per-element solution serves as the initial guess for the upcoming optimization procedure, thereby accelerating convergence and avoiding very poor local minima. Subsequently, a binary spatial mask is introduced to explicitly define the quiet zone, assigning a logical value of one to grid points within the spherical region satisfying Eq.~\eqref{eq:qz}, and zero elsewhere, thereby isolating the optimization domain and enhancing computational efficiency. During each iteration of the gradient-based optimization routine, individual RIS elements sequentially test incremental phase variations~$\Phi_{n} \pm \Delta\phi$, starting from an initial step size~$\Delta\phi$. For every candidate phase configuration~$\mathbf{\Phi}$, the average scattered field magnitude~$E_{\text{field}}^{\text{avg}}$ within the quiet zone is recalculated, and each RIS element is updated according to:
\begin{equation}
\Phi_{n} \leftarrow \arg\min_{\Phi_{n}\pm\Delta\phi} E_{\text{field}}^{\text{avg}}.
\end{equation}
After completing updates for all RIS elements, if no further improvement is detected, the step size~$\Delta\phi$ is halved to refine the search. The procedure iteratively continues until the power level within the quiet zone falls below a predefined threshold or the maximum number of iterations is reached, ultimately resulting in a minimized residual scattered field within the suppression region.

\subsubsection{Simulation Setup for Indoor Scenario} In this scenario,  we consider a cubic room with side length $L = 10~\text{m}$, where each RIS unit comprises $50 \times 50$ elements and is positioned along the walls with a margin of $m = 10\%$ from the edges of the respective surfaces. The operating frequency is set to $28~\text{GHz}$, reflecting typical sub-6~GHz indoor deployments scaled to mmWave bands. The objective is to establish as HSSA a spherical quiet zone at the center of the room, defined by coordinates $(L/2, L/2, 0)$ and radius $R_0 = 0.5  \text{m}$, while minimizing the impact on the FSDA that consists of all the surrounding regions. The environment outside the quiet zone is discretized into a three-dimensional grid with $N_x = N_y = N_z = 251$ points whereas within it the denser grid consists of about $3,400$ points for more accurate computations.  Regarding the optimization procedure, the maximum number of iterations is fixed at 150, with a convergence tolerance of $10^{-9}$.

Fig. \ref{fig:indoor_initial} illustrates the initial $\left| E_{\text{field}} \right|$ distribution within the room under the default RIS configuration, where both the source and the designated quiet zone are marked with red circles, and the field magnitude within the quiet zone is nearly identical to that of the surrounding space. This uniformity highlights the absence of any suppression mechanism, as the quiet zone initially experiences an average field magnitude of $2.01\text{V/m}$, closely matching the $2.14~\text{V/m}$ observed elsewhere. Following the application of SHIELD, the optimized distribution in Fig. \ref{fig:indoor_final} reveals a marked attenuation within the quiet zone, now visually and quantitatively distinguished from the rest of the environment. Specifically, the average field magnitude within the quiet zone is reduced to $4.5\times10^{-3}\text{V/m}$, while the surrounding area remains effectively unchanged at $2.13~\text{V/m}$, confirming SHIELD’s ability to deliver strong, localized suppression without disrupting the broader EM behavior. The extent and precision of this suppression are further quantified in Fig.~\ref{fig:difference_indoor}, which shows the point-wise deviation $\Delta E$ across the domain. Within the quiet zone, suppression reaches approximately $-49$~dB, while outside, deviations remain minimal, ranging from $-1$~dB in adjacent points to $+1.5$~dB at greater distances. This outcome directly reflects the spatial mask and optimization process, which are specifically tuned to isolate EM suppression to the intended region while preserving normal field propagation elsewhere. Moreover, the ability to achieve such deep suppression in a confined area, with negligible impact beyond its boundaries, demonstrates the potential of SHIELD as a practical tool for spatially controlled EM quiet zones.

\begin{figure}
\centering
\includegraphics[width=\linewidth]{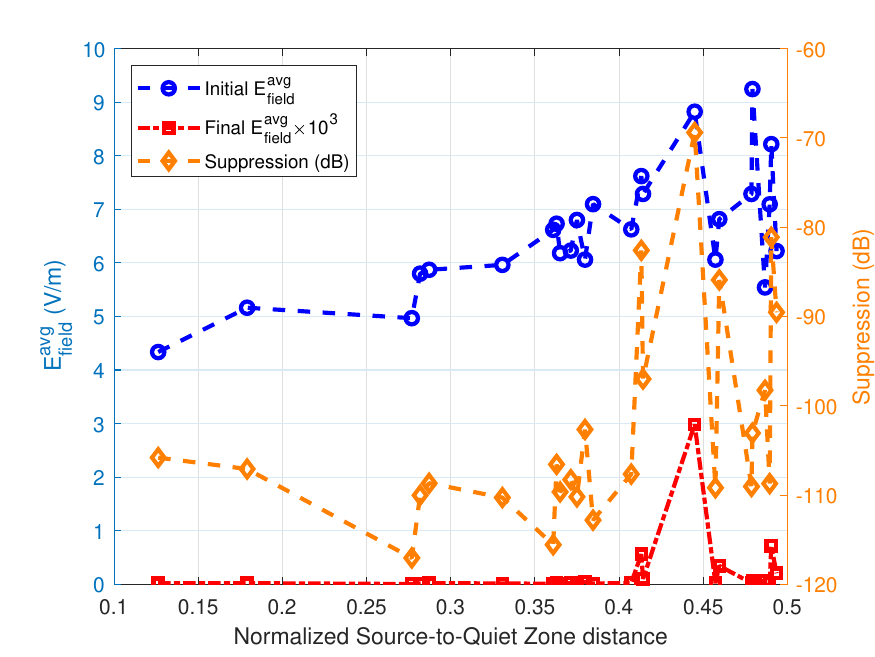}
\caption{Initial and optimized values of $E_{\text{field}}^{\text{avg}}$, along with the corresponding suppression levels.}\label{fig:indoor_results}
\end{figure}

Fig. \ref{fig:indoor_results} illustrates the initial and optimized values of the average electric field $E_{\text{field}}^{\text{avg}}$ within the quiet zone, together with the corresponding suppression levels in dB, plotted against the normalized distance between the source and the quiet zone relative to the room dimension $L$. The analysis considers multiple source positions along a circular trajectory centered at $(L/2, L/2, 0)$, ranging from a radius of 1 meter—coinciding with the quiet zone boundary—up to 5 meters, thereby encompassing diverse spatial configurations. Throughout this evaluation, the suppression region and computational settings remain fixed, ensuring that the results solely reflect the influence of source placement. Under default RIS conditions, the average $E_{\text{field}}^{\text{avg}}$ within the quiet zone remains consistently between $4.3$ and $9.2~\text{V/m}$ across all positions, highlighting the absence of any intrinsic spatial shielding. After optimization, the quiet zone field magnitude is reduced uniformly to approximately $2.2 \times 10^{-4}~\text{V/m}$, with a standard deviation of $6 \times 10^{-4}~\text{V/m}$, yielding suppression levels between $-69$ and $-117$~dB. Therefore, Fig. \ref{fig:indoor_results} confirms the capability of SHIELD to achieve stable and effective suppression, irrespective of source location, thereby reinforcing its suitability for consistent quiet-zone formation in dynamic scenarios.

\begin{figure*}[t]
  \centering
  \begin{subfigure}[b]{0.33\textwidth}
    \includegraphics[width=\linewidth]{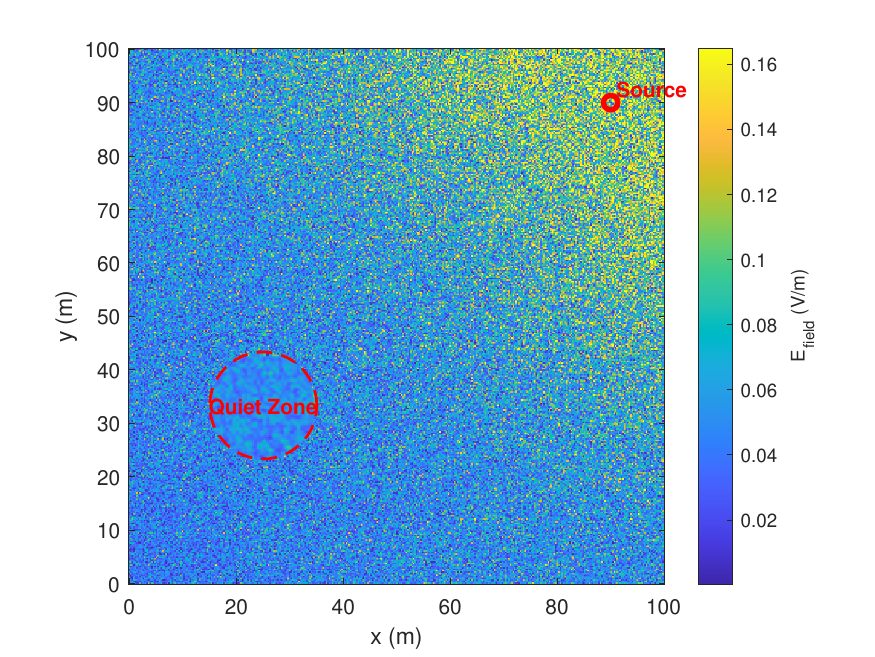}
    \caption{}
    \label{fig:outdoor_initial}
  \end{subfigure}\hfill%
  \begin{subfigure}[b]{0.33\textwidth}
    \includegraphics[width=\linewidth]{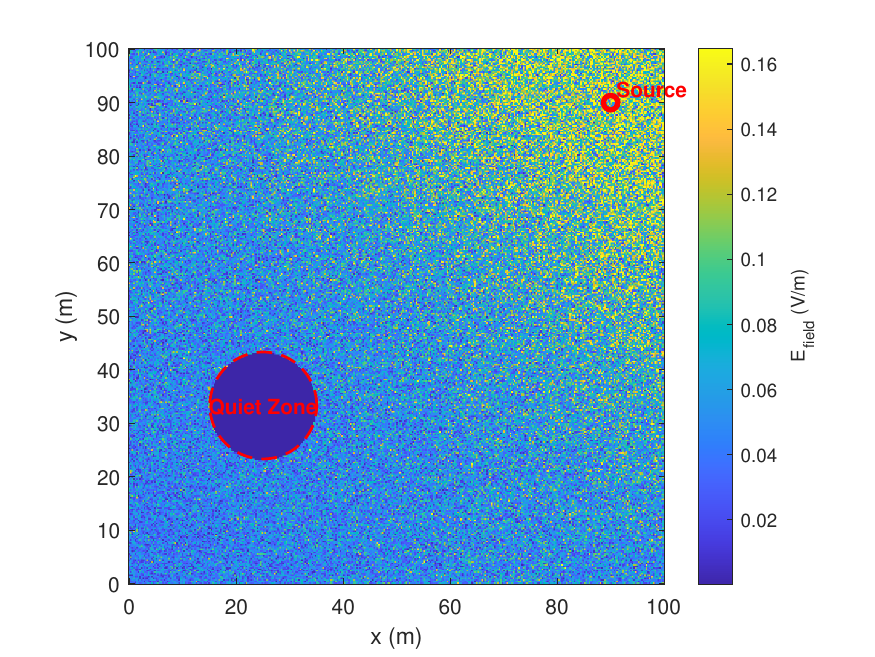}
    \caption{}
    \label{fig:oudoor_final}
  \end{subfigure}\hfill%
  \begin{subfigure}[b]{0.33\textwidth}
    \includegraphics[width=\linewidth]{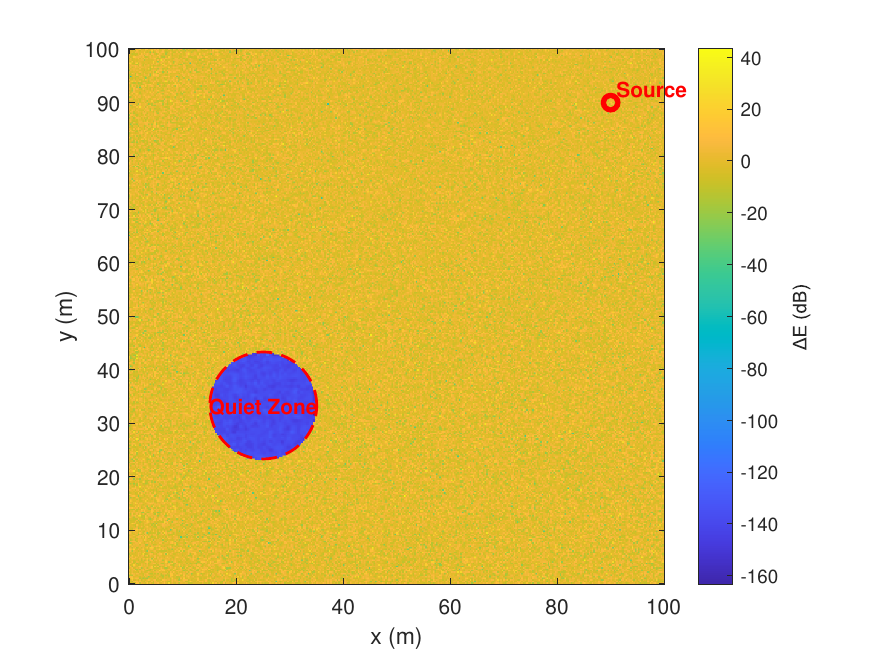}
    \caption{}
    \label{fig:difference_outdoor}
  \end{subfigure}
  \caption{Spatial distribution of $E_{\text{field}}$ (in V/m) before (left) and after (center) SHIELD, and the corresponding point-wise deviation (right) in dB.}
  \label{fig:outdoor}
\end{figure*}

\subsubsection{Simulation Setup for Outdoor Scenario} In this scenario, we consider the creation of a quiet zone within an open-area scenario with side length $L = 100$ meters. In this setup, RIS units are positioned on the surrounding buildings, each comprising $N_{\text{el}} = 150 \times 150$ elements to ensure comprehensive spatial coverage. The quiet zone is defined with a radius of $R_0 = 3$ meters, centered at coordinates $(L/4, L/3, 0)$, intentionally offset from the geometric center of the area to emulate realistic deployment scenarios. The operating frequency is maintained at $28$~ GHz. The discretization of the computational grid without the quiet zone is set to $N_x = N_y = N_z = 351$ whereas within it the grid consists of $22,400$ points for more precise results. The optimization process runs for 50 iterations, using a convergence tolerance of $10^{-9}$. 

Figs. \ref{fig:outdoor_initial} and \ref{fig:oudoor_final} illustrate the spatial distributions of $E_{\mathrm{field}}$ before and after the application of SHIELD, with the source located at $\left( \frac{9L}{10}, \frac{9L}{10}, 0 \right)$. Initially, the average scattered field magnitude within the quiet zone is approximately $5.1\times10^{-2}$~V/m, slightly lower than the $6.6\times10^{-2}$~V/m observed outside, primarily due to the greater distance from the source. After SHIELD is applied, the field within the quiet zone is suppressed to $6.5\times10^{-9}$~V/m, while the field outside remains stable at $6.3\times10^{-2}$V/m, demonstrating that suppression is confined to the designated region. This effect is further corroborated by Fig. \ref{fig:difference_outdoor}, where the deviation reaches approximately $-164$~dB within the quiet zone, and remains limited to the range of $-0.2$ to $+0.25$~dB elsewhere. These findings confirm SHIELD’s scalability and its ability to achieve highly localized suppression in outdoor scenarios, effectively managing RIS-induced reflections.

\section{Conclusion}\label{sec:conclusion}

In this paper, we introduced RF-Fencing, a novel RIS-enabled paradigm designed to enhance covert and secure communications within PWEs. To realize this vision, we developed SHIELD, a dedicated algorithm that dynamically orchestrates spatially selective suppression zones while simultaneously preserving high-quality connectivity in designated service areas. Through comprehensive simulations across both mmWave and THz frequency bands, we demonstrated SHIELD’s adaptability to diverse deployment conditions and its ability to effectively balance between suppression and service delivery, even under challenging scenarios with multiple closely spaced target regions. Beyond directional control, SHIELD also showcased its potential for establishing fully localized quiet zones, enabling practical EM isolation within indoor and outdoor spaces. These results collectively underline SHIELD’s promise as a scalable, computationally efficient solution for privacy-aware wireless control, paving the way toward secure and adaptive communication architectures in future 6G networks.

\bibliographystyle{IEEEtran}
\bibliography{refs}

\end{document}